\documentclass[prb,twocolumn,showpacs,preprintnumbers,amsmath,amssymb,superscriptaddress,unsortedaddress]{revtex4}

\usepackage{graphicx}
\usepackage{bm}
\usepackage{color}
\usepackage{amsmath}
\usepackage{amssymb}

\begin{document}

\title{Target bias voltage effect on properties of magnetic tunnel junctions by biased target ion beam deposition}
\author {Wei Chen}
\thanks{Corresponding author}
\email{wc6e@virginia.edu}
\affiliation{Department of Physics, University of Virginia, Charlottesville, VA 22904, USA}
\author {Dao N. H. Nam}
\affiliation{Department of Materials Science and Engineering, University of Virginia, Charlottesville, VA 22904, USA}
\affiliation{Institute of Materials Science, VAST, 18 Hoang-Quoc-Viet, Hanoi, Vietnam}
\author {Jiwei Lu}
\author {Kevin G. West}
\affiliation{Department of Materials Science and Engineering, University of Virginia, Charlottesville, VA 22904, USA}
\author {Stuart A. Wolf}
\affiliation{Department of Physics, University of Virginia, Charlottesville, VA 22904, USA}
\affiliation{Department of Materials Science and Engineering, University of Virginia, Charlottesville, VA 22904, USA}
\date{\today}
\begin{abstract}
Magnetic tunnel junctions (MTJ) with AlO$_x$ barrier were fabricated by a deposition tool called Biased Target Ion Beam Deposition (BTIBD) using low energy ion source (0-50 eV) and voltage biased targets. The BTIBD system applies bias voltage directly and only on the desired targets, providing enough sputtering energy and avoiding "overspill" contamination during film deposition. The successful deposition of AlO$_x$-MTJs demonstrated the capability of BTIBD to make multilayer structures with good film quality. MTJ thin film surface roughness and intermixing between layers are among the key problems leading to low TMR performance. Here by studying the bias voltage effect on MTJ properties via the measurement of N\'{e}el coupling field and TMR, we suggest that the lower bias voltage reduces the intermixing that occurs when top CoFe free layer is deposited on AlO$_x$ barrier, but produces relatively high surface roughness. On the other hand, higher energy deposition enhances both interlayer mixing and surface flattening. Such understanding of bias voltage effects on film properties could be used to optimize the MTJ performances.
\end{abstract}
%
%
\maketitle
\section{Introduction}
Magnetic tunnel junction (MTJ) has been a promising candidate for spintronic devices applications\cite{Wolf} such as magnetic random access memory and hard drive read head, and this is due to its large tunneling magnetoresistance (TMR) and good switching properties. It is well known that TMR properties are extremely sensitive to atomic scale interfacial roughness and interlayer mixing across multilayer interfaces, especially the interfaces with the tunnel barrier. Among many techniques to grow MTJ structures, ion beam deposition (IBD) has the advantages of low processing pressures, directional sputtered flux, high energy of the adatom flux, low sputtering rate and independent control of target and substrate environments.\cite{Hylton} However, conventional IBD methods were not designed for optimal deposition conditions for MTJ multilayer structures. For example, the conventional IBD can only be operated using relatively high sputtering ion energies and relatively low pressure with no substantial background gas scattering.\cite{Quan} This is because it is harder to focus the sputtering ion beam at low energies, meaning a larger fraction of the ion beam could miss the targets and sputter off undesired materials from the vacuum system hardware, causing overspill contamination.\cite{Hylton} As a result, the adatoms in IBD can have very high energies when they reach the substrate, causing severe intermixing problems at the interfaces, which is detrimental to high TMR performance.

The biased target ion beam deposition (BTIBD) system was developed to overcome some problems of the conventional IBD. The key advantage of the BTIBD system is the use of novel low energy ion sources that combine the end-Hall ion source\cite{Kaufman} and hollow cathode electron source,\cite{Kaufman2} which produce a very high density of inert gas ions with a very low energy (from several eV). By applying a negative bias to the desired target, only the biased target materials will be sputtered off for deposition by the accelerated high energy ions, while the portion of ions missing the target will not cause any damage to the environment due to the un-accelerated low energy for which the sputter yield of the typical vacuum system hardware is negligible. The broad low energy ion beam combined with the easily controlled target biasing (50-1200 eV) gives much freedom in terms of interface engineering, making it a well-suited tool for depositing multilayer structures like MTJ.

In this paper, the capability of the BTIBD system is demonstrated by depositing standard MTJ with naturally oxidized AlO$_x$ barrier. It is well known that for MTJ thin film stacks with layer thickness on the order of nanometers, the film surface roughness and intermixing between adjacent layers are among the key problems  not only preventing the successful switching of the free layer, but also resulting in reduced TMR and unfavored Resistance-Area product (RA) values. Others have studied the effects of deposition rate on properties of single film\cite{Vopsaroiu} and multilayers\cite{Manzoor,Schrag} to understand the dependence of film morphology on deposition rate, in hopes of reducing surface roughness and interlayer mixing for optimized growth conditions. Here, with a different approach, the effects of bias voltage (i.e. different deposition energy) on MTJ properties is studied via the investigation of the offset field of the free layer due to the N\'{e}el coupling and the TMR performance, and an optimization strategy based on the study to reduce both film surface roughness and intermixing is proposed. This study also demonstrates the BTIBD's  capability of precise thin film and interface engineering at the atomic level.

\section{Experiment}
\begin{figure}[t!]
  \includegraphics[width=8.0cm]{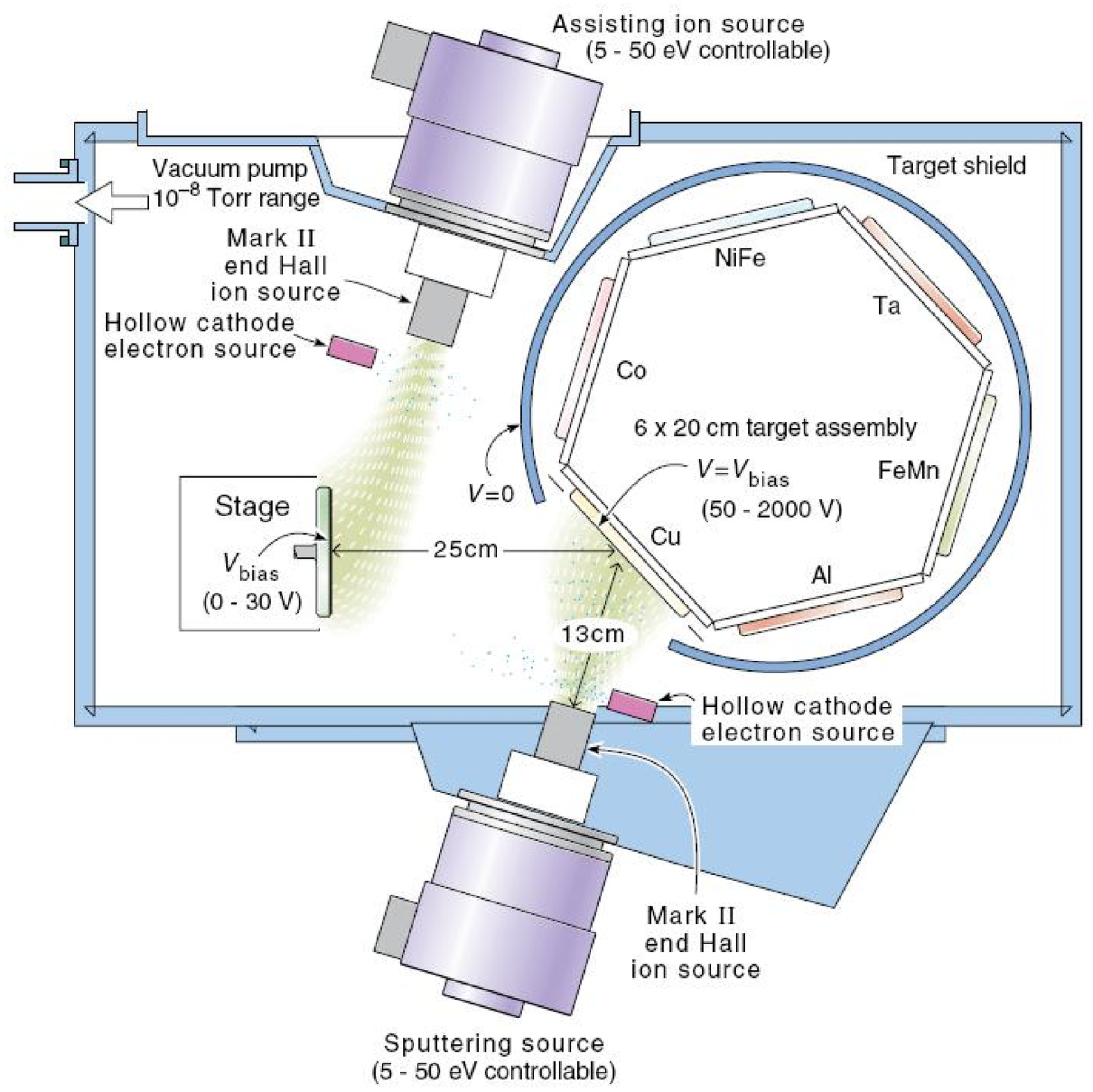}
  \caption{Schematic illustration of the BTIBD system.}\label{Fig1}
\end{figure}

The schematic diagram of the BTIBD used in this work is shown here in Fig. \ref{Fig1} for the reader's convenience. Additional details of our BTIBD system can also be found elsewhere.\cite{Hylton,Zhurin} There are two identical ion guns installed in the BTIBD, one for main deposition and the other for ion beam assisted deposition. In this paper, only the main gun was used for MTJ deposition to simplify the initial study.

MTJ samples were deposited on clean thermally oxidized Si substrates, with the typical structure Ta(5)/CoFe(3)/FeMn(6)/CoFe(3)/AlO$_x$(2)/CoFe(3)/Ta(5) [units in nm]. The base pressure of the BTIBD system was around 5$\times$10$^{-8}$ Torr, and the working pressure was close to 7$\times$10$^{-4}$ Torr during the deposition. A 50 Oe field was applied during the film deposition of CoFe and FeMn layer to create the magnetic easy axis, and a natural oxidation method was taken to make the AlO$_x$ barrier. Oxidation was carried out in multiple steps to produce better quality oxide barrier.\cite{Chen} For instance, a nominal thickness of 5 {\AA} Al was deposited and naturally oxidized. This process was repeated four times to get a 2 nm AlO$_x$ barrier. The first 5 {\AA} Al layer was oxidized in 500 mTorr of O$_2$ for 5 minutes and the the next three Al layers were oxidized in 1 Torr for 10 minutes. To study the effect of bias voltage on MTJ properties, different bias voltages are applied when depositing the the top CoFe free layer. Deposition rates were calculated from film thicknesses measured by atomic force microscopy (AFM), and the deposition time was adjusted to make the top CoFe free layer with the same thickness for various target bias voltages. To create unidirectional exchange anisotropy, all the samples were quickly field-cooled from 250 $^\mathrm{o}$C in 3 kOe external field in the forming gas environment. We measured the magnetic properties using a Quantum Design PPMS-6000, and obtained the TMR data by current-in-plane tunneling technique (CIPTech)\cite{Worledge} measurements of the un-patterned blanket films. In this study, two sets of samples were made to investigate the target voltage effect on MTJ properties. The first set consisted of three samples with the same typical structure mentioned previously. All of the layers of the three samples were deposited with 600 V bias voltage except for each of the top CoFe free layers, where bias voltages 300 V, 600 V and 900 V were used, respectively. For the second set, all conditions were the same as the first except the thickness of the middle CoFe layer. Instead of 3 nm used in the first set, we increased it to be 4 nm for all three samples in the second set.

\begin{figure}[b!]
  \includegraphics[width=8.5cm]{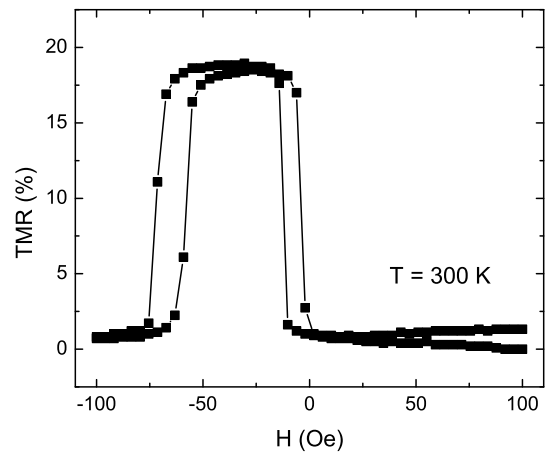}
  \caption{Representative TMR result from the CIPTech measurement.}\label{Fig2}
\end{figure}

\section{Results and Discussion}
Fig. \ref{Fig2} shows the simulated TMR value from the CIPTech measurement of a MTJ sample with nominal pre-oxidized Al layer thickness of 2 nm, which is the second sample in the second set, with all the layers deposited with a target bias voltage of 600 V and middle CoFe layer thickness 4 nm. The TMR value of this sample is close to 20\%. This is not a record high value, since it is not our objective to optimize the TMR value. Additionally, our CoFe target has a composition of Co$_{95}$Fe$_5$, which was found not to be the optimal composition for producing high TMR.\cite{Yang}

\begin{figure}[b!]
  \includegraphics[width=8.5cm]{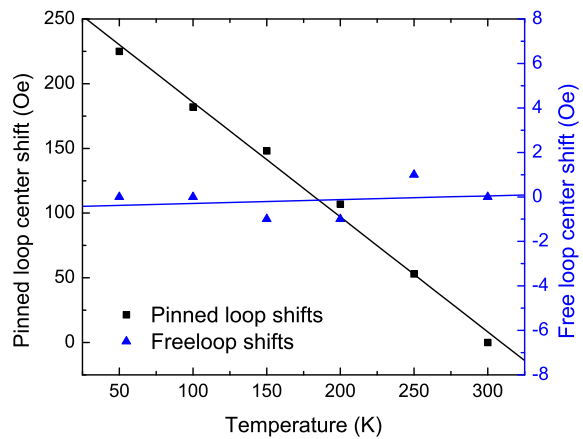}
  \caption{Loop center shifts of both free and pinned layers with different temperatures of the sample with 600V free layer bias and 4nm middle CoFe layer.}\label{Fig3}
\end{figure}

As far as the interlayer magnetic coupling between the two ferromagnetic electrodes of MTJ is concerned, previous studies have shown that two separate effects tend to produce extraneous magnetic fields in the plane of free layer: magnetostatic coupling due to uncompensated poles near the sample edges and N\'{e}el coupling due to interfacial roughness.\cite{Schrag} For un-patterned samples in our case, the magnetostatic coupling is negligible. Another requirement before we can assume that N\'{e}el coupling is the only magnetic coupling affecting the free layer switching is that the tunnel barrier of the MTJ must be pinhole free. Without this condition, we cannot separate the indirect N\'{e}el coupling from the direct magnetic coupling through the pinholes. To prove that the MTJ is pinhole free, we measured the temperature dependence of the hysteresis loop center shifts of both the free and pinned layers of all the MTJ samples, and compared the loop center shifts at different temperatures to the value at 300 K. Fig. \ref{Fig3} shows a representative measurement of the second sample in second set with free layer deposited with 600 V and 4 nm middle CoFe layer. In the Figure, we see the increasing loop center shifts of the pinned layer as the temperature decreases. This is because the pinning strength of the antiferromagnetic FeMn gets larger at lower temperature (i.e., the exchange bias of the pinned layer increases as the temperature decreases). However, the loop center shifts of the free layer remain almost constant according to the temperature change. If there were pinholes, the free layer would be directly coupled with the pinned layer. Then, as the pinned layer loop shifts with temperature, the direct magnetic coupling would force the free layer loop to shift as well. This result suggests that, according to Pong \textit{et al.}\cite{Pong} the AlO$_x$ barrier in this sample is free of pinholes. Similar results indicating pinhole-free-barriers were obtained for the other five samples in our investigation.

\begin{figure}[t!]
  \includegraphics[width=8.5cm]{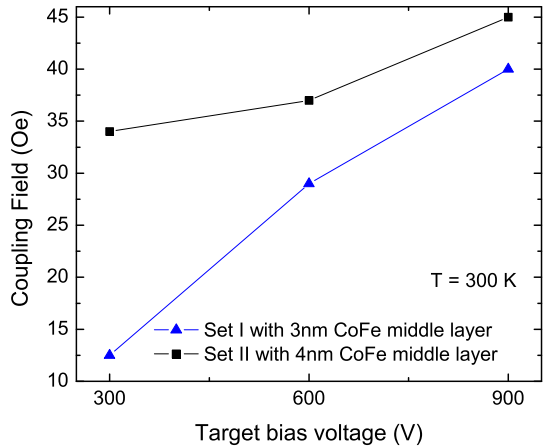}
  \caption{The dependence of free layer loop center shift on the target bias voltage used for the free layer deposition.}\label{Fig4}
\end{figure}
According to N\'{e}el's model,\cite{Neel,Schrag} N\'{e}el coupling is closely related to the film and interface morphology. In this model, a sinusoidal roughness profile is assumed, and the N\'{e}el coupling field is given by
\begin{equation}
H_{N}=\frac{\pi^{2}}{\sqrt{2}}(\frac{h^{2}}{\lambda t_{F}})M_{P}e^{(-2\pi{\sqrt{2}}t_{s}/\lambda)},
\label{eqno1}
\end{equation}where $\lambda$ and \textit{h} are the amplitude and wavelength of the roughness profile, \textit{t$_F$} and \textit{t$_s$} are the thickness of the free layer and that of the barrier, and \textit{M$_P$} is the magnetization of the pinned layer.

Since N\'{e}el coupling has been proven to be the only effect responsible for the free layer hysteresis loop center shift in our samples, we can extract the information about the interfacial mixing and film surface roughness by studying the N\'{e}el coupling field of the free layer. Fig. \ref{Fig4} shows the free layer loop center shifts at room temperature for the two sets of samples with details described at the end of Section II. First, we examine the results of the loop center shifts between the two sets. The second set, with thicker middle CoFe layer (i.e. thicker pinned ferromagnetic layer with larger M$_P$), always shows higher N\'{e}el coupling offset field of the free layer when compared to individual samples with the same bias voltage in the first set. This is qualitatively consistent with N\'{e}el's model presented in Eq. (1). Additionally, the Figure shows that within each sample set, the N\'{e}el coupling field is stronger with increased target bias voltage for the free layer deposition. There are two possible competing factors when considering the impact of high energy adatoms on film morphology. One is that the high energy adatom of the free layer, when hitting the barrier surface, could redistribute the surface clusters and flatten the barrier surface, effectively increasing the barrier thickness, so the coupling field would be reduced according to N\'{e}el's model. In a opposite way, adatoms with high energy would cause significant interlayer mixing between the top free CoFe layer and the barrier layer underneath, which effectively makes a rougher interface compared to the barrier surface before the free layer deposition. This leads to a stronger N\'{e}el coupling and therefore a larger loop center shift of the free layer hysteresis loop. From the above result in Fig. \ref{Fig4}, the stronger N\'{e}el coupling with increased bias voltage indicates at least the existence of an intermixing effect and it should be dominant even if the barrier surface is flattened during free layer deposition.

\begin{figure}[t!]
  \includegraphics[width=8.5cm]{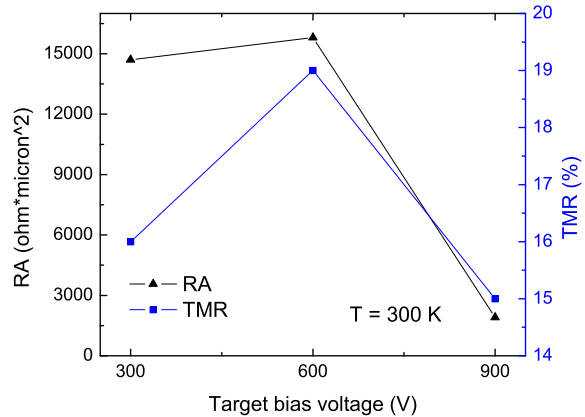}
  \caption{The dependence of TMR and RA values on the target bias voltage used for the free layer deposition.}\label{Fig5}
\end{figure}

To better study the two possible factors mentioned above due to bias voltage effect, we also investigated the TMR and RA values dependence on bias voltage. Fig. \ref{Fig5} shows the TMR and RA values of the second sample set with different bias voltages applied for free layer deposition. As we just explained, the higher energy causes more interlayer mixing between the tunnel barrier and free layer. This intermixing could smear the thin barrier, causing more electron scattering at the interface, so lower TMR value would be expected. Also such intermixing can effectively make the barrier thinner, so the RA value would decrease with higher bias voltage too. On the other hand, the possible barrier surface flattening, due to impact of the same high energy adatom at the beginning stage of deposition, could effectively make the barrier thicker and sharper, so we would expect larger RA and TMR values. As shown in the Figure, both RA and TMR values are maximized at 600 V compared with the values at 300 V and 900 V. Such result indicates that both the factors of intermixing and surface flattening coexist and compete with each other. From 300 V to 600 V, the initial barrier surface flattening due to higher adatom energy helps to get higher RA and TMR values. While form 600 V to 900 V, the high adatom energy makes the interlayer mixing the dominant effect, causing both reduced RA and TMR values. The strategy based on this study would be as follows: for maximum TMR value, the optimized bias voltage should be some intermediate level depending on the material properties. For applications that require minimized N\'{e}el coupling, a low bias voltage would be preferred.

This study focuses mainly on the impact of bias voltage on the interfacial properties. As to the film surface morphology, previous molecular dynamics simulation work\cite{Zhou} showed that higher adatom energy would result in a smoother film surface. Three single CoFe film samples were deposited with same thickness but different bias voltages of 300 V, 600 V, and 900 V. The corresponding AFM measurement of the surface roughness generated the RMS value of 1.269 nm, 0.943 nm and 0.713 nm respectively, indicating the smoother sample surface with higher bias voltage. This result is consistent with the conclusion from the simulation work.

\section{Conclusion}
MTJs with AlO$_x$ barrier were deposited using the BTIBD sputtering tool. The application of low energy source and target voltage bias makes it well suited for multilayer film deposition like MTJ, which requires atomic level interfacial engineering. The study of bias voltage effects on the interfacial and film surface properties via N\'{e}el coupling facilitated our understanding of the adatom energy impact on the film properties, and the strategy based on our study for the best performance of MTJs would be using high bias voltage for the bottom ferromagnetic electrode deposition to get a smooth seeding surface, and using optimized intermediate bias level for the free layer to get high TMR value. For applications that require minimized N\'{e}el coupling, the low bias voltage would be preferred.

\begin{acknowledgments}
Special thanks to Dr. W. F. Egelhoff, Jr. and Dr. P. J. Chen in NIST for helpful discussion and the CIPTech measurements. This work was carried out under the support of the Defense Microelectronics Activity (DMEA) through a subcontract with University of California, Riverside.
\end{acknowledgments}

\end{document}